\documentclass[aps,prl,superscriptaddress,showpacs,endfloats,preprint]{revtex4}
\usepackage{graphicx}

\begin{document}

\title{Constraints induced by finite plasma formation time on some physical observables at RHIC}
\author{V.~S.~Pantuev\\
University at Stony Brook, Stony Brook, NY 11794-3800
}
\date{\today}
\begin{abstract}
We discuss consequences of finite plasma formation time at RHIC. 
Nuclear modification factor, azimuthal assymetry, di-jet correlations can quantitatively be described 
by particle production in the early stage of the nuclear collision. A possible impact on interpretation 
of the non-photonic electrons and J/$\psi$ data is also considered.

\end{abstract}



\maketitle

The existence of a long Quark Gluon Plasma (QGP) formation time~\cite{paper}, T=2-3 fm/c, 
may have a strong inpact on some physical observables 
in the experiment. Direct physical interpretation of these observables could be missleading. 
If jets  or other hard scattering processes do not suffer from strong energy loss or absorption 
during formation time, there should be a significant corona effect. All processes in the corona region will not be influenced 
by the produced dense medium. The relative contribution of corona  decreases with centrality, but 
could account for at least 20$\%$ of 
hard processes in the most central collisions. 
It means, for example, that at each centrality class, the nuclear modification factor $R_{AA}$ for a particle 
produced in a hard scattering proccess can't be smaller than 
the corona contribution. Of caurse, this is valid if an additional absorption, like normal nuclear absorption, 
does not takes place during these first 2-3 fm/c. 
Many physical observables will be influenced by contribution from the corona region. 
A list of possible cosequences follows. 
We used a simple  Monte Carlo simulation 
of nucleus-nucleus collisions based on Glauber approach with a Woods-Saxon nuclear density distribution~\cite{paper}.\\

\begin{figure*}[hbt]
\begin{minipage}[t]{0.8\linewidth}
\includegraphics[width=1.0\linewidth]{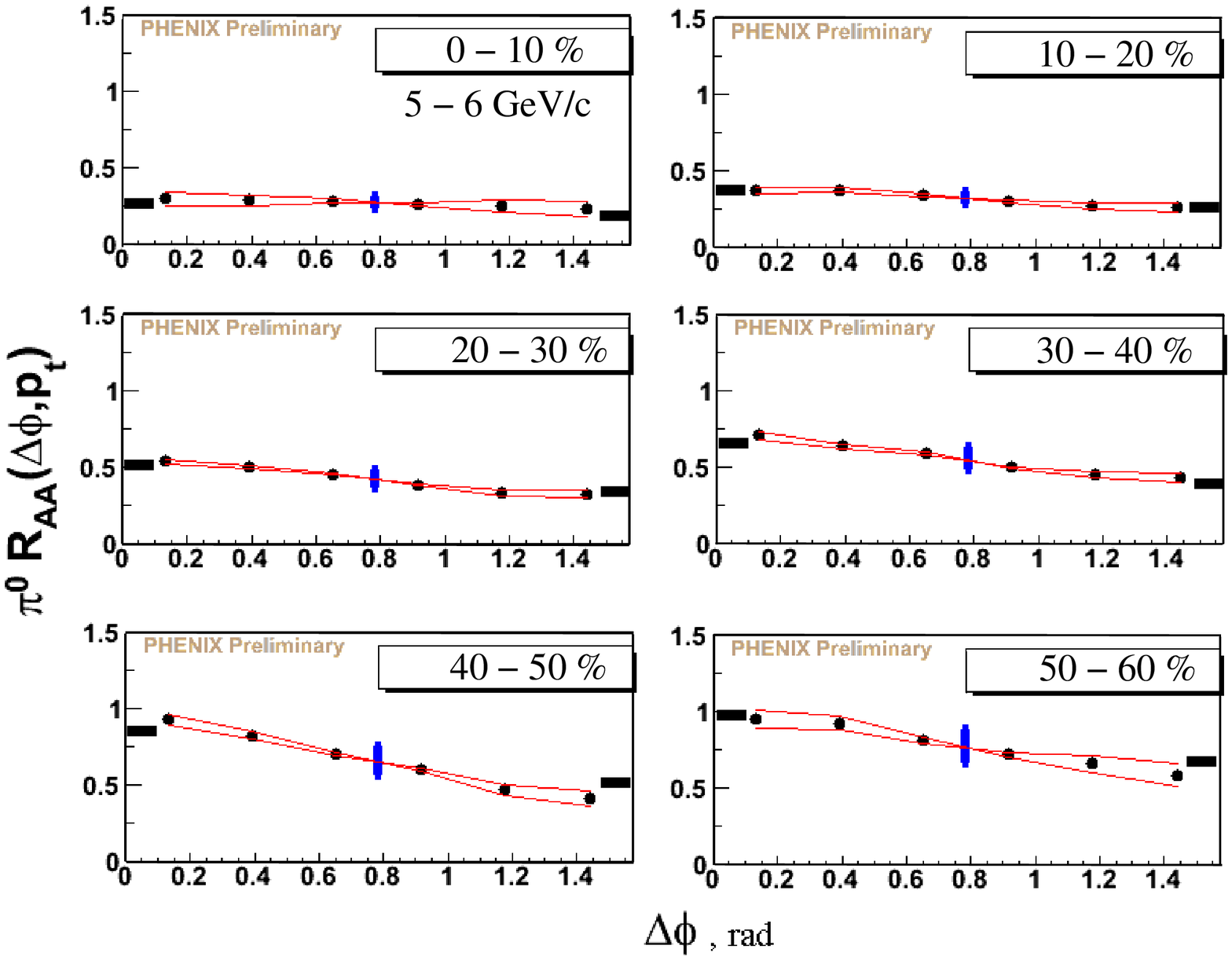}
\end{minipage}
\caption{\label{fig:Raa_vs_phi} PHENIX preliminary  $R_{AA}$ results for $\pi^0$ at momenta 5-6 GeV/c 
versus angle $\phi$ relative 
to the reaction plane for different centralities in Au+Au collisions at RHIC~\cite{winter}. Points are experimental data, 
thin lines show systematic errors from the reaction plain resolution and vertical bars in the middle show 
the reaction plane  avaraged $R_{AA}$ value and its error.
Black horizontal bars are predictions of our model.}
\end{figure*}

1. The PHENIX collaboration at Relativistic Heavy Ion Collider, RHIC, 
showed a preliminary result for the $R_{AA}$ dependence on azimuthal angle relative to the reaction plane in Au+Au 
collisions at $\sqrt{s_{NN}}$=200 GeV at RHIC ~\cite{david, winter}, Fig.~\ref{fig:Raa_vs_phi}.
As expected,  $R_{AA}$ is larger in the reaction plane and smaller out of the plane. 
The most interesting feature of this result is that in the event centrality class 50-60$\%$,  
the in-plane $R_{AA}$ becomes 
close to one. That is, no absorption at all is seen for high $p_t$ pions. At the same time a significant particle 
absorption is seen in the out-of-plane $R_{AA}$. 
Here $R_{AA}$ is a nuclear modification factor which is defined as 
\begin{equation}
R_{AA}(p_T) 
= \frac{(1/N_{evt}) \; d^{2}N^{A+A}/dp_T d\eta }
{(\langle N_{binary} \rangle/\sigma^{N+N}_{inel}) \; d^{2}\sigma^{N+N}/dp_T d\eta},
\label{eq:RAA_defined}
\end{equation}
where $\langle N_{binary} \rangle$ is a number of binary nucleon-nucleon collisions at a particular 
centrality class.
In the 50-60$\%$ event centrality class,  
the amount of nuclear matter is still significant in all directions. It's puzzling that a high momentum pion 
can ``punch through'' the interaction zone in the in-plane direction but may be stopped when traveling in the other direction. 
In the paper~\cite{paper} we attempted to explain numerically this feature by a QGP formation time of $T=2.3 fm/c$. During
this time, for certrality class 50-60$\%$, in-plane produced jets can freely leave the interaction zone. From other side, 
jets produced out-plane, where the size of interaction region is larger, have not enough time to escape and will be partially absorbed by the 
formed dense medium.

2. Obvious consequences of such a scenario are: a) all pions above 5-6 GeV/c are produced by parton fragmentation 
from the 
corona region, which is formed between 2 to 3 fm/c after the beginning of the collision; b) combining this with 
the experimental fact that the hadron to pion ratio, enhanced at low momenta, returns back to its vacuum value 
at $p_t$ above 5 GeV/c~\cite{flatRaa}, we can conclude that $all$ high $p_t$ hadrons are produced from corona region; 
c) $R_{AA}$ versus $p_t$ for pions and hadrons at 
high transverse momenta will be flat~\cite{flatRaa}.

\begin{figure}[thb]
\includegraphics[width=1.0\linewidth]{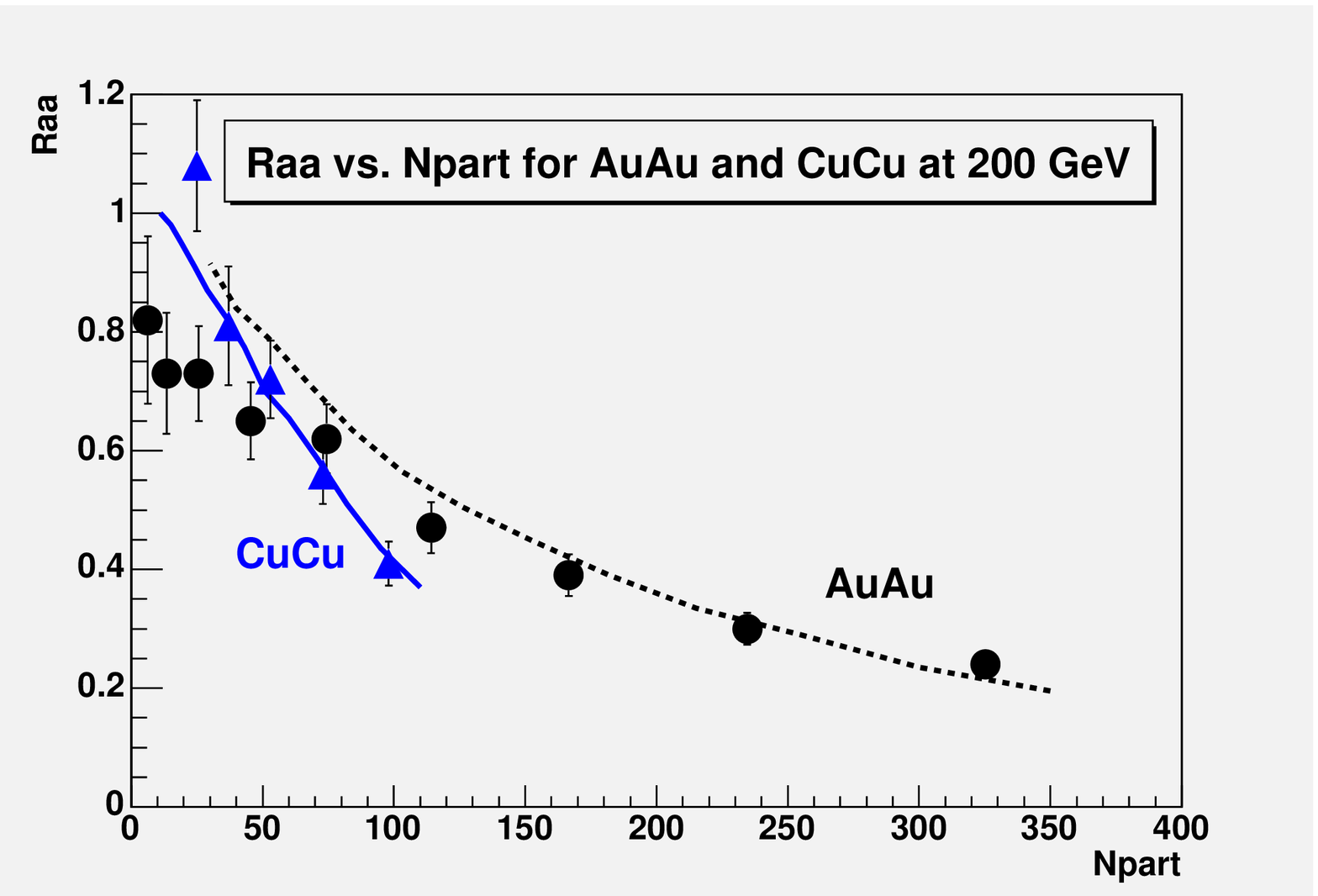}
\caption{\label{fig:Cu}  Calculated $R_{AA}$ for Au+Au, dashed curve, and Cu+Cu, solid curve, collisons versus the number of
participant nucleons, $N_{part}$. The circles are experimental data for
$\pi^0$ yield for Au+Au intergrated for $p_t$$\ge$4 GeV/c~\cite{flatRaa}. The triangles are data for Cu+Cu intergrated for 
$p_t$$\ge$7 GeV/c~\cite{maya}.}
\end{figure}

3. The model parameter $T=2.3 fm/c$ was adjusted to describe a small subset of Au+Au data at 
centrality  50-60$\%$. However, with this value of $T$ we can nicely explain the $R_{AA}$ behaviour 
not only for all centrality classes of Au+Au collisions, but for Cu+Cu interactions too, Fig.~\ref{fig:Cu}.  
%

\begin{figure}[thb]
\includegraphics[width=1.0\linewidth]{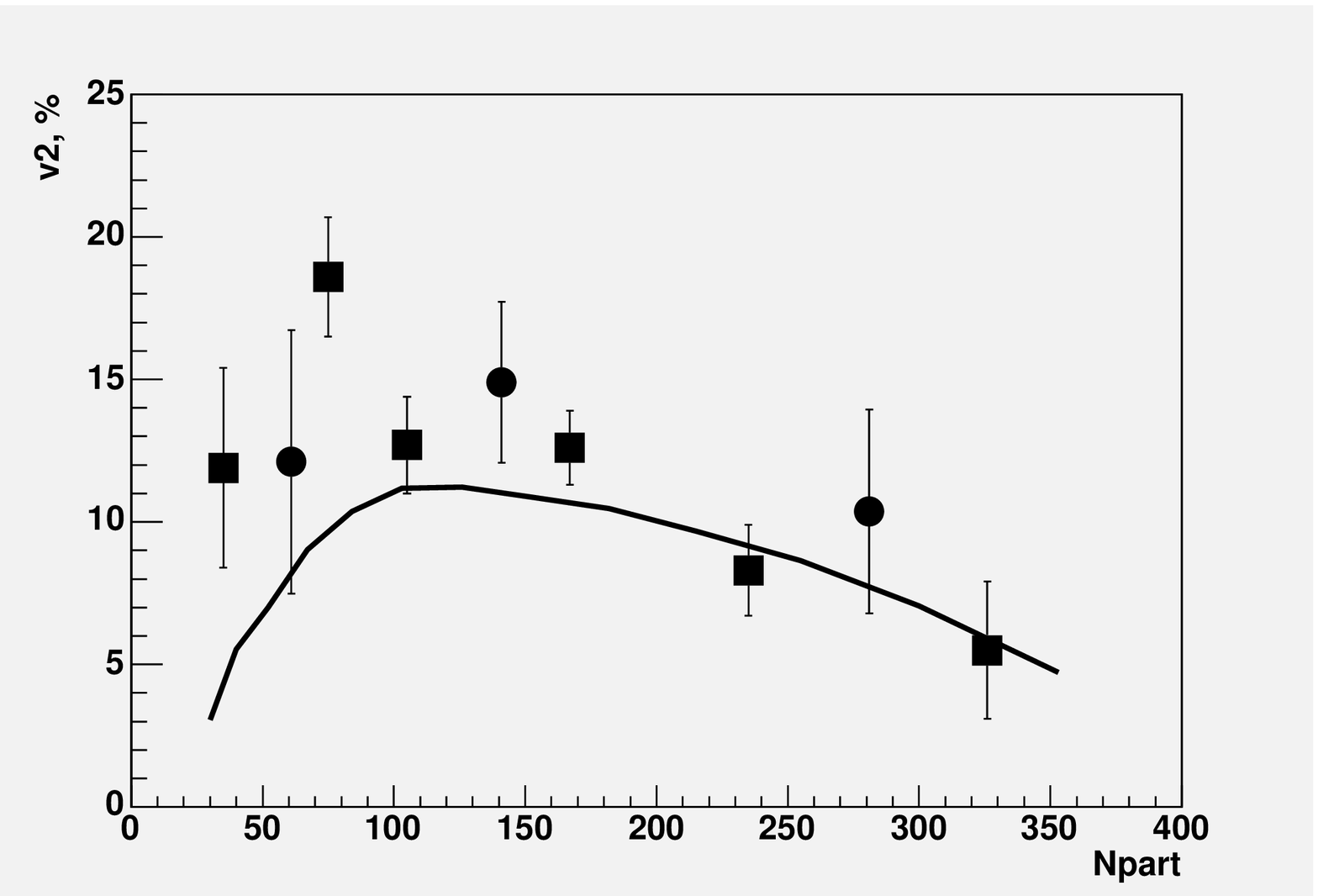}
\caption{\label{fig:v2} Calculated ellipticity parameter $v_2$ for Au+Au collisons, solid line, versus the number of
participant nucleons, $N_{part}$. Data for $\pi^0$ with error bars are: squares for 5-7 GeV/c, cicles for 4.59 GeV/c. 
PHENIX preliminary data~\cite{kaneta, winter}.}
\end{figure}

4. We also estimate another experimental variable, $v_2$. It describes the elliptic shape of the nucleus-nucleus
collision in azimuthal angle as  dN/d$\phi$=N(1+2$v_2$cos(2$\phi$)). The long formation time naturally brings 
the result of non-zero $v_2$ for high $p_t$ pions~\cite{paper}, Fig.~\ref{fig:v2}. The value of $v_2$ reaches 11-12$\%$ 
in mid-central 
collisions of gold nuclei with Woods-Saxon  density distributions. At each centrality class $v_2$ 
will be constant at high transverse momenta. 
%

5. Two-particle angular correlations of hadrons  close to the high $p_t$ ``trigger'' particle or  
opposite to the ``trigger'' particle directions allow to investigate  
of the near-side and the away-side jets properties, respectively. Following the argumentation that all particles at  high $p_t$ 
are produced from corona region, we can explain the weak dependence of properties of the near-side jets~\cite{jetshape} 
with centrality in Au+Au collisions. In the cited paper, some increase  
was observed in the near-side particle yield only for low momentum  associated particles.

\begin{figure}[thb]
\includegraphics[width=1.0\linewidth]{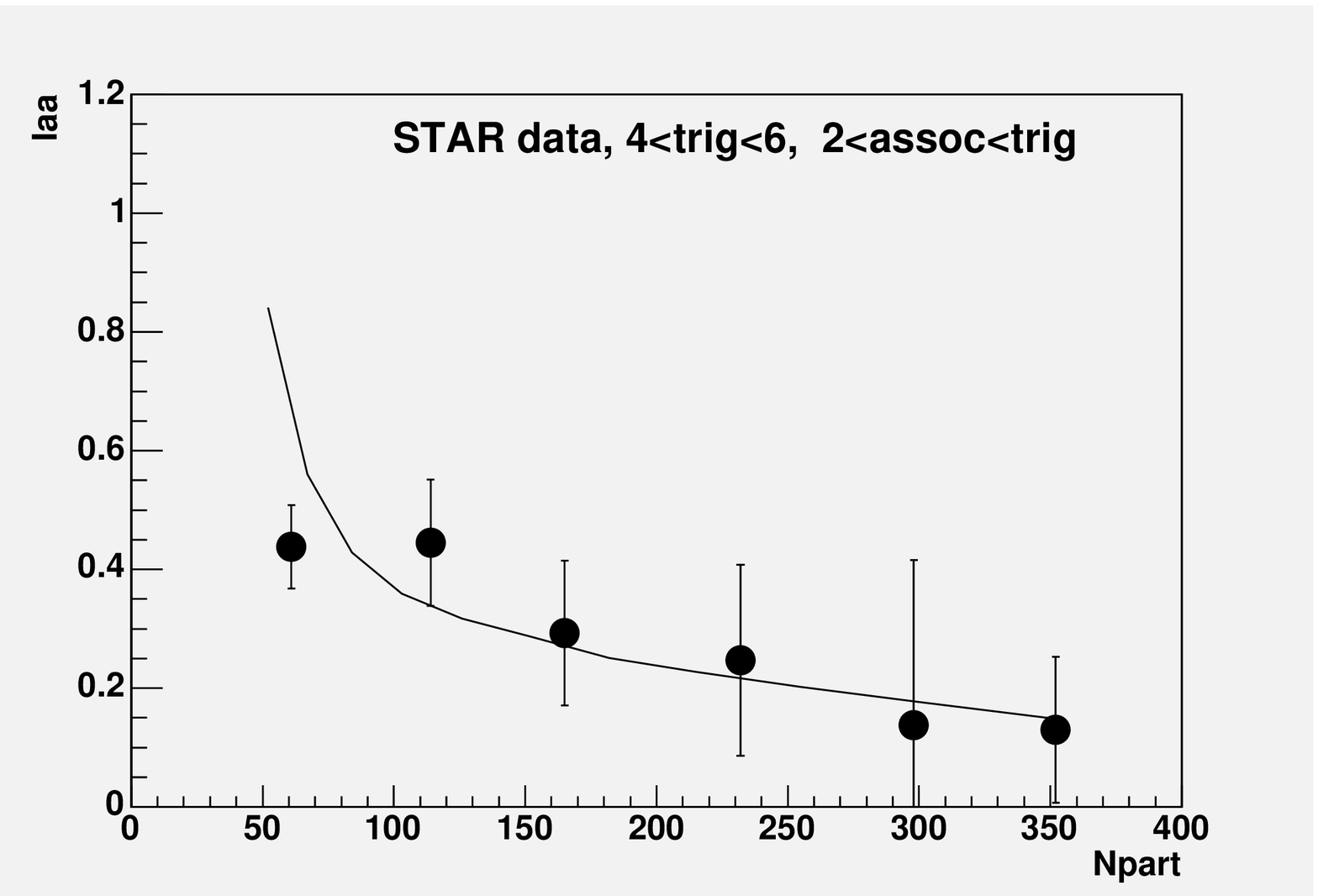}
\caption{\label{fig:Iaa}  Calculated $I_{AA}$ for Au+Au, the curve, versus the number of
participant nucleons, $N_{part}$. The width (sigma) of away-side jet is 0.35 radians.
The cicles are experimental data from~\cite{dijets}. }
\end{figure}

6. STAR collaboration published the results on $I_{AA}$ in Au+Au collisions - the ratio of away-side yield per 
trigger particle to the similar value from p+p collisions~\cite{dijets}. We calculate $I_{AA}$ within our model 
with $T=2.3 fm/c$ for different centralities. The increase of corona thickness by 2.3 fm keeps the value of 
$I_{AA}$ at the level close to $R_{AA}$, Fig.~\ref{fig:Iaa}. We want to emphasize that these are two $different$ variables, 
$I_{AA}$ is 
related to the yield $per$ $trigger$ particle. The trigger particle should ``survive'' from absorption in a first 
place. In this calculation we assume that away-side jet distribution has a Gaussian shape with the width parameter of 0.35 radians.
%

\begin{figure}[thb]
\includegraphics[width=1.0\linewidth]{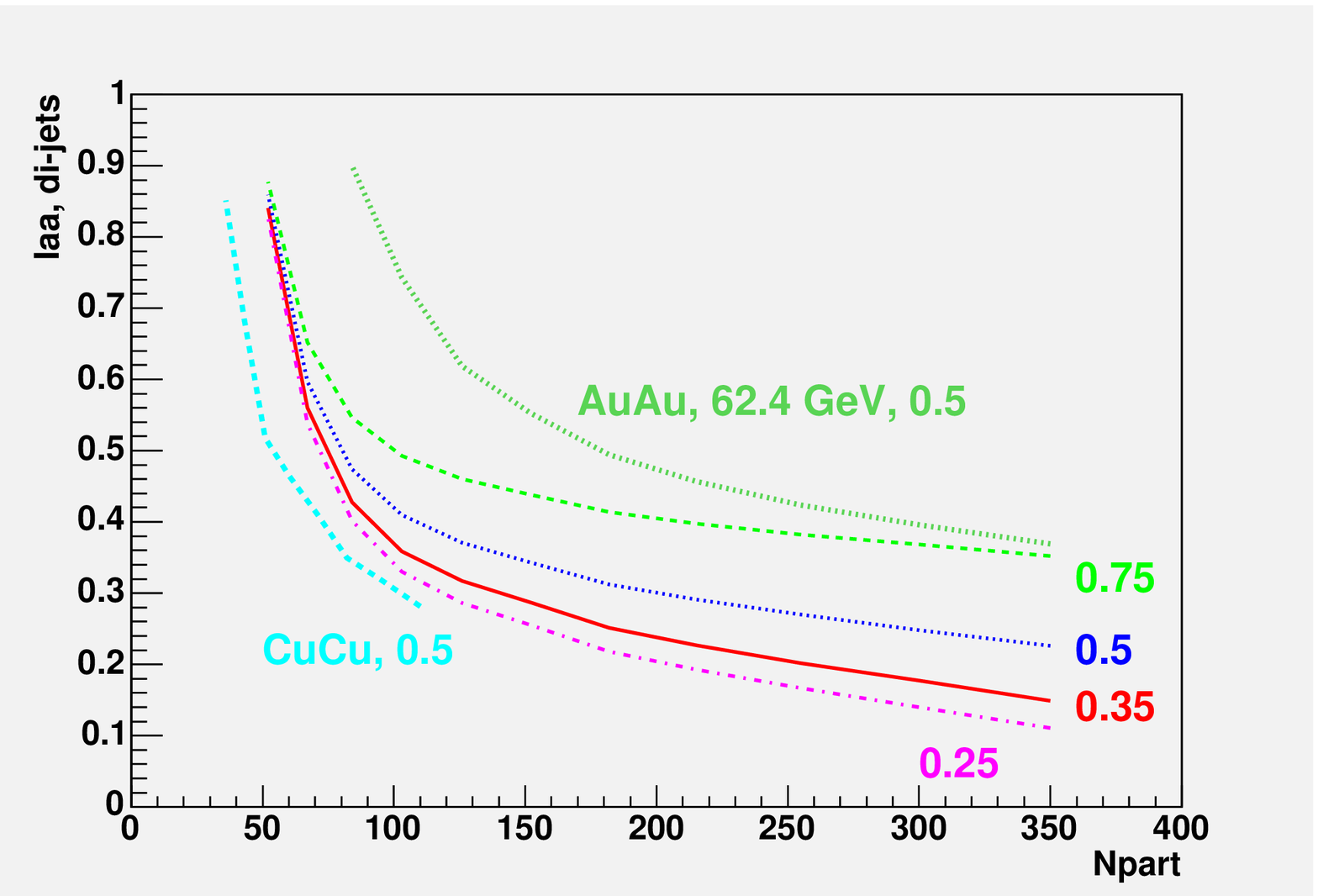}
\caption{\label{fig:Iaa_cuts}  Calculated $I_{AA}$ for Au+Au versus the number of
participant nucleons, $N_{part}$, for different width of away-side jet - the numbers in radians next to the lines. 
Results for Cu+Cu at 200 GeV and Au+Au collisions at 62.4 GeV are also shown.}
\end{figure}

7. At Quark Matter 2005 conference, August 4-9, in Budapest, Hungary, STAR collaboration claims that 
away-side jet $re-appears$ at high  trigger and associated particles momenta~\cite{newdijets}.
While changing the momentum range of the associated particle, the jet width will change too - for higher
momentum particles the jet width should be smaller.   
We check the STAR result by changing the  width of simulated away-side jet, Fig.~\ref{fig:Iaa_cuts}.
For all values of width, $I_{AA}$ is still significant even for most central events. We argue that 
in previous analysis at lower momentum range, the away-side jet was wider and it was not easy to reconstract 
its remnants on the level of 20-25$\%$. Besides this, at low associated particle momentum, the responce of the high 
density medium on the absorbed away-side jet changes dramatically in shape and in yield~\cite{cone}. 

\begin{figure}[thb]
\includegraphics[width=1.0\linewidth]{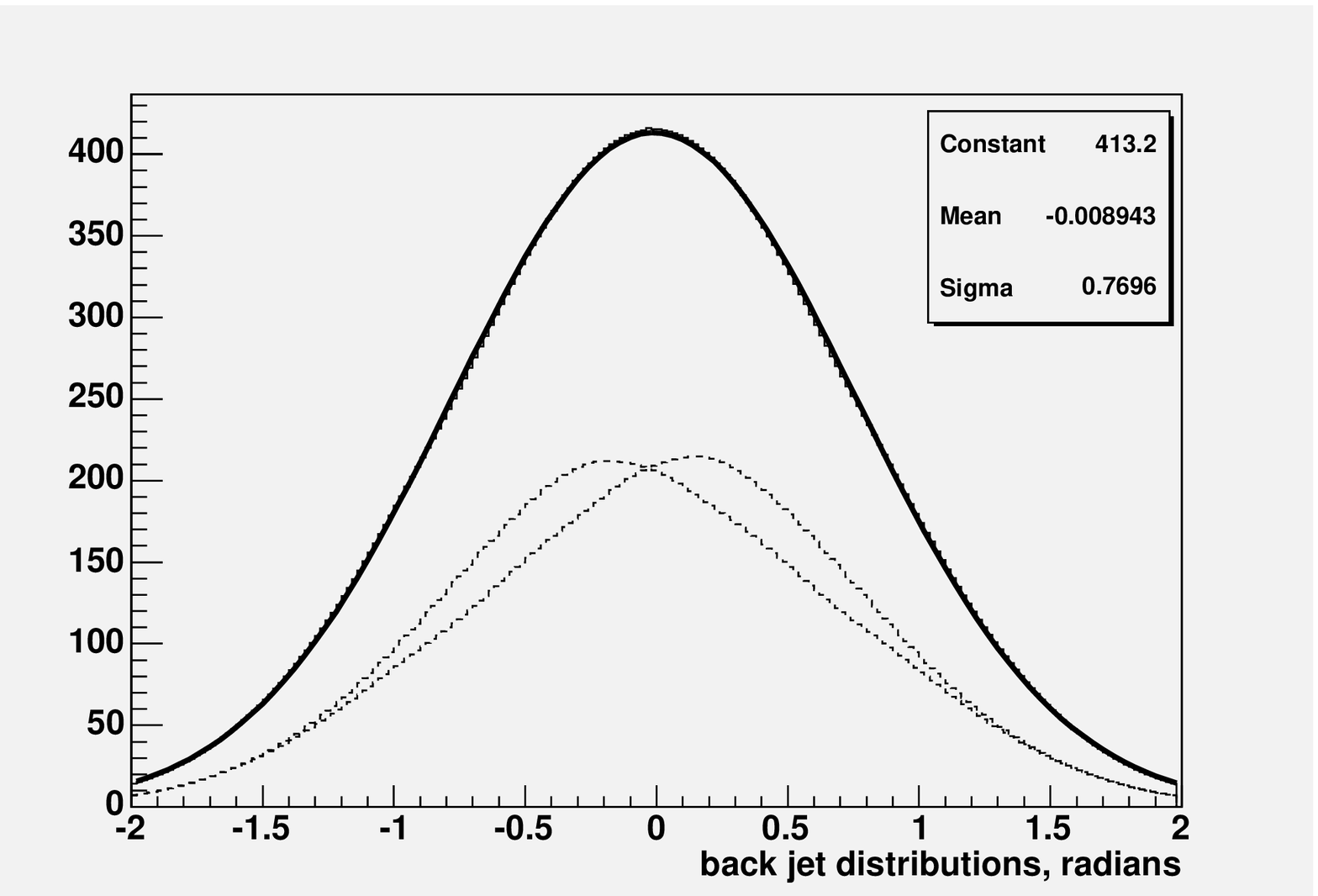}
\caption{\label{fig:jetshape} Angular distributions within the away-side jet in the case of jet production from 
the corona region with strong jet absorption
in QGP core. The original away-side jet had a Gaussian shape with the width parameter of 0.75 radians.
Dashed histograms represent the jet distribution from two different sides of the collision zone.
The solid line histogram is the sum of these two subsets. Numbers in the box represent the result of the 
solid line histogram fit by the Gaussian function.
}
\end{figure}

8. Discussions in the previous parts 2, 6 and 7 lead to the conclusion that, probably, even at particle 
momenta 5-10 GeV/c, we don't see ``punch-through'' jets, these are di-jets from corona region. This is why 
width of the away-side jet does not change with centrality and is the same as in p+p collisions~\cite{newdijets}. 
We investigate the away-side jet shape in our model. Di-jet forms kind of a tangential emission from the 
surface of  
interaction region. If in experiment the trigger particle radiates at particular direction, di-jets can 
survive along the two tangents, from the edges of the collision zone. If the collision zone 
is small and away-jet is wide we may see jet splitting too. In Fig.~\ref{fig:jetshape} we plot the 
results of the simulation of di-jets in mid-central 30-35$\%$ Au+Au events. The away-side jet originally had a 
Gaussian shape with the width of 0.75 radians. The dashed histograms represent jets survived at 
the two tangential directions. We see slight shifts from the original direction on the order of 25$\%$ of the original 
width. 
The solid line histogram 
represents the sum of these two subsets and its fit by the Gaussian function. The width parameter 
sigma of this fit, 
 0.77 radians, is very close to the original width. The observed features do not change significantly with 
centrality. We can conclude that, indeed, the shape and the width of away-side jet are not changed much. 
There is no visible back jet splitting produced in the corona region, thus it can not be an explanation 
of the two peak structures seen in the experiment~\cite{cone} at low momentum of the associate particle. 
%

\begin{figure}[thb]
\includegraphics[width=1.0\linewidth]{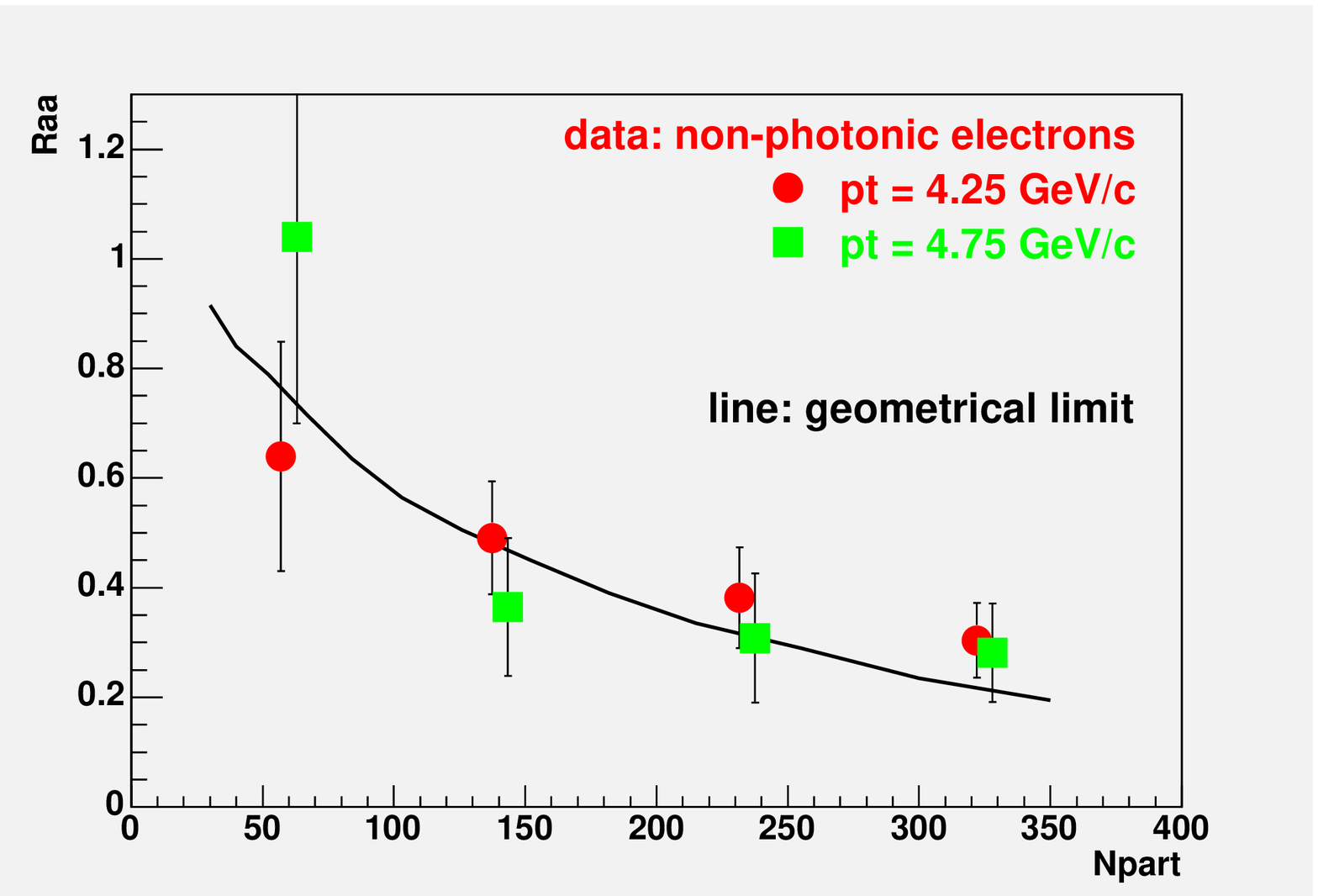}
\caption{\label{fig:charm} $R_{AA}$ for  non-photonic electrons in Au+Au versus the number of
participant nucleons, $N_{part}$. The PHENIX preliminary experimental data are from~\cite{charm}: circles and squares are for 
$p_t$=4.25 GeV/c and 4.75 GeV/c, respectively. Shown are statistical errors only. 
Solid line - the result of our calculation which puts the geometrical limit in case of strong charm quark absorption.}
\end{figure}

9. In Fig.~\ref{fig:charm} we present PHENIX preliminary results on non-photonic electrons at high momenta 
shown at Quark Matter 2005~\cite{charm}. We also plot the lower limit of $R_{AA}$ determined by 
particle production from the  corona region during the formation time, solid line. Experimental points sit exactly on 
the line. These electrons originated from D-mesons, charmed baryons and, probably, from 
bottom quark decays. The conclusion of Fig.~\ref{fig:charm} would be that at high $p_t$ the charm quarks are completely 
absorbed by
dense medium, similar to the light quarks.
%

10. In the case of a strong charm quark absorption, following the argument in our previous part 4, one can 
expect a significant $v_2$ for non-photonic electrons. Experiments indeed found non-zero $v_2$~\cite{charm, laue}. 
Qualitativety, it is possble to descibe also the momentum dependence of $v_2$. Leptonic modes of D-meson decay 
are mostly
with 3- and 4-particles in the final state. The low momentum electrons should be produced isotropically from 
the low energy D-mesons. With increasing energy of the meson, decay products will start to follow the direction of 
decaying meson, so electron $v_2$ will increase too
(this argument was also stressed in the reference~\cite{laue}), and finally will saturate for high $p_t$
at the values presented in 
Fig.~\ref{fig:v2}. 
It would be mistaken to say naively ``charm flows''. This statment needs clearefication and an additional 
experimental investigation.

\begin{figure}[thb]
\includegraphics[width=1.0\linewidth]{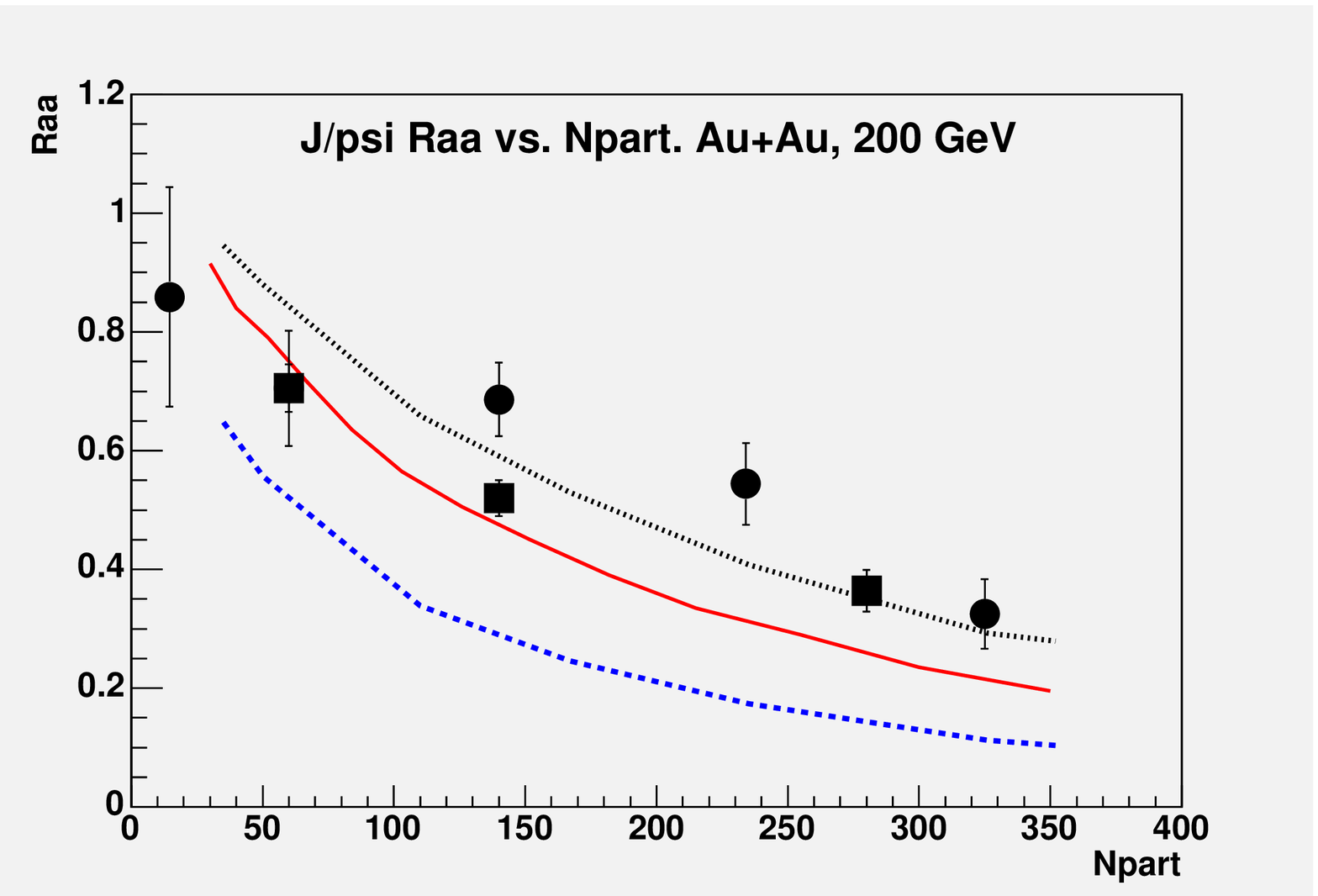}``gray''
\caption{\label{fig:Jpsi} $R_{AA}$ for J/$\psi$ in Au+Au versus the number of
participant nucleons, $N_{part}$. The PHENIX preliminary experimental data are from~\cite{Jpsi}: 
cicles - at midrapidity, squares - at rapidity $y=1.2-2.2$. Shown are statistical errors only. 
For explanations of the different lines, see discussion in the text.}
\end{figure}

11. We expect many interesting features from J/$\psi$ production in nuceus-nucleus collisions. J/$\psi$ is a
closed system of two heavy charm quarks. In particular interest is the interation of such a system with the surrounding 
medium. Previously, a very strong J/$\psi$ suppression was observed at lower energy nucleus-nucleus 
collisions~\cite{na50, na50b}. 
New PHENIX preliminary data show the suppression at about similar strength~\cite{Jpsi},  Fig.~\ref{fig:Jpsi}.
In our consideration we devide nucleus-nucleus collision in to two stages: before plasma formation and 
after. As it was already demonstrated, hard scattering processes, like jet production, do not suffer from 
energy loss or absorption during plasma formation time. The matter is transparent for them at this stage.
In contrast, during the actual ``plasma time'', the medium looks completely ``black'' for jets.\\
Part of J/$\psi$s could be absorbed during early stage of the collision. 
In this case the medium will be ``grey'' for them. At the same time, other models predict incomplete 
suppression of J/$\psi$ in QGP phase, which means also that plasma is ``grey''. We try to demonstrate how
the distinction of  plasma formation time, QGP itself and different suppression processes lead to a few major
scenario, Fig.~\ref{fig:Jpsi}.  Scenario (a), $\textit{white and black}$: no absorption at the early stage, 
total absorption 
in the QGP - solid line in Fig.~\ref{fig:Jpsi}. It this case everything is similar to the jet absorption. 
Scenario (b), $\textit{grey and black}$: 
during the plasma 
formation time J/$\psi$ has normal nuclear absorption but suffers from total absorption in the QGP, lower dashed line. This line 
goes obviously below the solid curve and was obtained by applying the results of R.Vogt calculations~\cite{vogt} 
to the corona region only. In other words, it is the solid line result from the case (a) times the original curve from R.Vogt. 
This would be an overestimation of the normal nuclear absorption because we did not take into account that the initial 
stage lasts 2.3 fm/c only. Scenario (c), $\textit{white and grey}$: top dotted line, where we assume that there is 
no absorption during the initial stage and some suppression in the QGP stage. Here we take just one model with QGP 
absorption by Kostyuk et al.~\cite{kostyuk} and applied to the second, QGP, phase only.\\
It is worthwhile to mention that the original calculations~\cite{vogt, kostyuk} for J/$\psi$ we used in cases (b) and (c) significantly 
overestimate and underestimate $R_{AA}$, respectively~\cite{Jpsi}. \\
By listing such naive speculations we want  to stress the importance of the different stages 
of the nuclear collision, including plasma formation time, for J/$\psi$ production.

In conclusion, we want to emphasize that we propose not ``yet another'' simple model with just
one parameter, which actually, may increase a little bit if the longitudinal expantion will be taken into 
account. 
We do not pretend to describe in many details all features of the 
nucleus-nucleus collision at RHIC. We present an $orthogonal$ point of view on some experimental 
data. This view could be useful for better understanding of properties of a new matter produced 
at RHIC. Perhapes, it will bring clear vision or some new ideas on formation process and properties 
of a quark-gluon phase.

We tried to explain some of the features seen in the experiment, but also bring two major questions:
where this formation time comes from, why models with fractional parton energy loss do not work well 
or why energy loss is very strong? The answers may come soon: interesting new approaches such as the complex 
plasma model~\cite{thoma} or the 
radically new picture of QCD with polymer chains have been proposed~\cite{shuryak} recently.



\begin{references}
\bibitem{xyscan} For all experimental data with the stamp ``preliminary'' we scaned avalible electronic 
version of the plots by program $xyscan$ developed by Thomas S. Ullrich.
\end{references}
\end{document}